\documentclass[a4paper,10pt]{article}
\usepackage[utf8]{inputenc}
\usepackage{amsmath}
\usepackage{graphicx}
\usepackage{authblk}
\usepackage{xcolor}
\usepackage{blindtext}
\usepackage[numbers,sort&compress]{natbib}
\usepackage{doi}
\usepackage{hyperref}
\hypersetup{
    colorlinks,
    linkcolor={red!50!black},
    citecolor={blue!50!black},
    urlcolor={blue!80!black}
}

\title{Laterally confined photonic crystal surface emitting laser based on monolayer tungsten disulfide operating at room temperature}
\author[1]{Xiaochen Ge}
\author[2]{Momchil Minkov}
\author[2]{Shanhui Fan}
\author[3]{Xiuling Li}
\author[1,*]{Weidong Zhou}
\affil[1]{University of Texas at Arlington, Arlington, TX 76019, USA}
\affil[2]{Stanford University, Stanford, CA 94305, USA}
\affil[3]{University of Illinois Urbana-Champaign, Urbana, IL 61801, USA}
\affil[*]{\href{mailto:wzhou@uta.edu}{\texttt{wzhou@uta.edu}}}
\date{}

\begin{document}

\maketitle

\begin{abstract}
We report a photonic crystal surface emitting laser using monolayer tungsten disulfide as the gain medium. The cavity design utilizes a heterostructure in the photonic crystal lattice to provide lateral confinement for a high quality factor with a compact active region. Room temperature continuous wave lasing is realized after integrating monolayer tungsten disulfide flakes onto the silicon nitride photonic crystal on quartz substrate. Highly directional, near surface normal emission has also been experimentally demonstrated.
\end{abstract}


As a direct band semiconductor, monolayer transition metal dichalcogenide (TMDC) draws tremendous research interests and numerous monolayer or few layer TMDC based lasers\cite{ye2015monolayer,wu2015monolayer,salehzadeh2015optically,li2017room,shang2017room,fang1305,wang2018monolayer,ge2018large} and light-emitting diodes\cite{withers2015light,liu2016nanocavity,yang2016electrically,bie2017mote,lien2018large} have emerged recently. With an ultimately thin active region, TMDC lasers have the potential of realizing ultra low lasing threshold and power consumption. The flexibility also enables integration possibilities on unconventional substrates\cite{lee2013flexible,de2016high}. The first demonstrations of TMDC lasers are based on microdisk cavities\cite{ye2015monolayer,salehzadeh2015optically} and photonic crystal (PhC) defect mode microcavites\cite{wu2015monolayer}. The emission wavelength is then extended to near infrared by introducing molybdenum ditelluride (MoTe$_2$) as the gain medium\cite{li2017room,fang1305}. While offering a low threshold and high Purcell factor, the emissions from microcavities typically have a large divergence angle, which limits their application in practical devices due to requirement of high numerical aperture (NA) lenses for collecting the emission. On the other hand, surface emitting devices have been implemented by optimizing the far field pattern of the PhC microcavity\cite{fang1305} and introducing distributed Bragg reflector (DBR)\cite{shang2017room} or plasmonic grating\cite{wang2018monolayer} based cavities.

Based on the Fano resonance at the $\Gamma$ point of the photonic band diagram\cite{fan2002analysis,zhou2014progress}, photonic crystal surface-emitting laser (PCSEL) is a promising laser cavity design as it provides simultaneously large area lasing and a high beam quality\cite{hirose2014watt,noda2017photonic}. Additionally, low index contrast cavity is also possible due to the relaxed index contrast requirement in such cavities. Such a unique advantage motivates its implementation in various gain mediums and a wide wavelength range\cite{chua2011low,baumann2007organic,zhao2016printed,liu2017photonic,kodigala2017lasing,matsubara2008gan}. However, although possessing a superior optical and electrical properties than the synthesized counterpart, the achievable size and yield of exfoliated monolayer TMDC flakes is still limited\cite{desai2016gold}, which makes TMDC based PCSEL challenging. For shrinking the size of the active region while maintaining the advantage of PCSEL, a lateral confinement scheme with heterostructure PhC was proposed, where modulation of the hole radii creates a mode gap near the $\Gamma$ point above the light cone\cite{ge2018low}. Here we will present the PCSEL design for monolayer tungsten disulfide (WS$_2$) based on such a principle. 

\begin{figure}[htb]
 \centering
 \includegraphics[width=\textwidth,keepaspectratio=true]{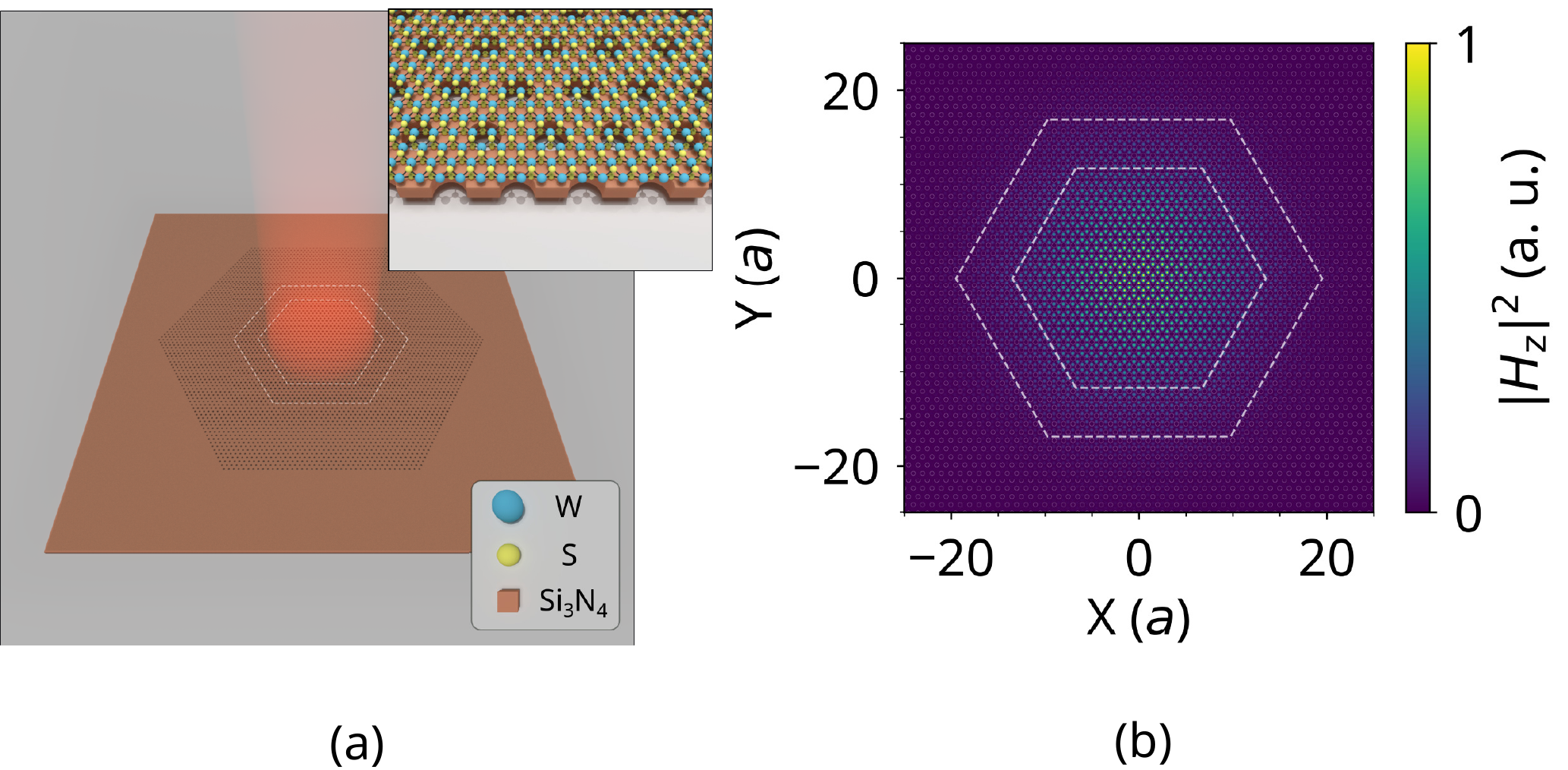}
 \caption{Design of the WS$_2$ based PCSEL. (a) Sketch of the cavity configuration and an illustration of the surface normal emission from the active region. The inset shows a cross-sectional view of the WS$_2$ integration scheme. (b) 3D FDTD simulated field intensity of the fundamental cavity mode near the cavity core region. In both subfigures white dashed lines are drawn at the interfaces of each PhC regions as a visual guide.}
 \label{fig:design}
\end{figure}

Fig.~\ref{fig:design}(a) shows a sketch of the cavity design, where a hexagonal PhC lattice of air holes on a $112~\text{nm}$ thick silicon nitride (Si$_3$N$_4$) slab on quartz, with the lattice constant $a = 454~\text{nm}$, is considered. The PhC lattice is composed of three regions with slightly different air hole radii, as indicated by the dashed lines. The center part of the lattice, denoted as the cavity core region, containing $15$ periods of air holes starting from the origin, is the active area of the laser cavity where the optical mode is localized. The lateral confinement is provided by the outer area of the PhC lattice with reduced hole radius compared to the core region, which is denoted as the cladding. And $6$ periods of air holes are placed between the core and cladding regions for a smooth transition of the hole radius, which improves the quality ($Q$) factor of the cavity mode. More specifically, the air hole radius in each PhC region is modulated to be $R_\text{core} = 0.25a$, $R_\text{trans} = 0.24a$ and $R_\text{clad} = 0.22a$, respectively. The cavity resonances are simulated by the finite-difference time-domain (FDTD) method\cite{oskooi2010meep}. The $H_\text{z}$ intensity of the fundamental TE-like mode, which has the highest $Q$, is plotted in Fig.~\ref{fig:design}(b) and the mode is well confined within the core region of the PhC cavity even though there is an absence of a complete in-plane photonic band gap. Vertically the mode is confined due to the existence of a bound state in continuum (BIC) at $\Gamma$\cite{ge2018low}. The theoretical resonant wavelength and $Q$ factor are determined to be $\lambda = 628.3~\text{nm}$ and $Q \approx 23,500$. The high $Q$ factor is crucial for realizing TMDC lasers as the monolayer gain medium results in a fundamentally limited confinement factor, which raises the material gain requirement for achieving lasing. 
As the PhC cavity mode is near the $\Gamma$ point in the band diagram, directional surface normal emission would be expected, as illustrated in Fig.~\ref{fig:design}(a). The inset of Fig.~\ref{fig:design}(a) shows the zoomed-in cross-sectional view of the schematic of monolayer WS$_2$ integration, where the flake is directly transferred onto the Si$_3$N$_4$ PhC, which can be done by the widely used transfer-printing processes\cite{castellanos2014deterministic}. 

For fabricating the PhC cavities, stoichiometric Si$_3$N$_4$ was deposited on quartz by low pressure chemical vapor deposited (LPCVD). The PhC cavities were patterned on ZEP520A e-beam resist by electron beam lithography (EBL) and transferred to Si$_3$N$_4$ by inductively coupled plasma reactive ion etching (ICP-RIE) using a gas combination of SF$_6$, CHF$_3$ and Helium. After stripping the resist in hot N-Methyl-2-pyrrolidone the sample was treated with oxygen plasma to remove residues. Fig.~\ref{fig:exp}(a) shows a scanning electron microscopy (SEM) image of a representative cavity sample near the core region.

Monolayer WS$_2$ was prepared by a gold-assisted exfoliation process\cite{desai2016gold}. Bulk WS$_2$ was firstly dispensed on tape. A layer of gold (100 nm) was deposited on WS$_2$ by electron beam evaporation. To exfoliate WS$_2$ flakes from bulk crystals, a polydimethylsiloxane (PDMS) stamp attached on a glass slide was used. Owing to the affinity between sulfur and gold\cite{magda2015exfoliation,desai2016gold}, large monolayer flakes can be obtained, which show profound color contrast under optical microscope inspection.  After being identified, the gold-coated WS$_2$ flakes were transfer-printed onto silicon oxide substrate using an automatic assembly platform (FiconTEC FL-200). The SiO$_2$ surface was then cleaned by oxygen plasma for activation and contamination removal. The gold layer was removed in KI/I$_2$ solution to expose the WS$_2$ flakes.

To transfer the WS$_2$ flakes onto Si$_3$N$_4$ PhCs, a layer of poly(methyl methacrylate) (PMMA) was coated on the SiO$_2$ as carrier and another piece of PDMS stamp was attached on the PMMA to provide mechanical support. The whole stack was submerged in deionized water. After applying ultrasonication, PMMA and SiO$_2$ were separated due to water penetration and the WS$_2$ flakes were also released from SiO$_2$ in the process. The WS$_2$ flakes, together with the PMMA carrier were aligned and transfer-printed onto the target Si$_3$N$_4$ PhC cavities. With the support of the PMMA film, the monolayer flakes can be completely transferred onto patterned substrate. Finally, the PMMA carrier was removed in acetone. Fig.~\ref{fig:exp}(b) shows the microscopy image of a fabricated device where the PhC cavity is completely covered by monolayer WS$_2$.

The fabricated sample was characterized on a micro-photoluminescence ($\mu$-PL) setup sketched in Fig.~\ref{fig:exp}(c). A $445~\text{nm}$ continuous-wave laser was used to pump the devices at room temperature. Light from the laser source was spatially filtered by a pinhole of $10~\mu\text{m}$ diameter and focused onto the sample surface by an objective lens of $0.25~\text{NA}$. The laser beam spot, when located at an unpatterned area, was captured by the camera and measured to be $11.77~\mu\text{m}^2$ using the second moment method.
The PL signal was collected by the same objective lens and directed to a monochromator (Horiba iHR-550) with a cooled charge-coupled device (CCD) camera (Horiba Synapse BIUV) for spectrum acquisition. A variable iris diaphragm with the opening diameter of $D = 2~\text{mm}$ was placed at the momentum space to restrict the angle of collection to $\theta = 3.2^\circ$. Prior to the actual measurement, the collection light path was aligned such that the emission signal normal to the sample surface would pass through the center of the iris diaphragm while the signal collected by the monochromator is maximized.

The measured PL spectra are plotted in Fig.~\ref{fig:exp}(d). The on-cavity spectrum was collected at the center of the cavity core region and the off-cavity spectrum was collected from the same monolayer WS$_2$ flake but outside of the PhC area. A sharp resonant peak corresponding to the fundamental cavity mode can be observed at $638.5~\text{nm}$ with the fitted full width half maximum linewidth of $0.27~\text{nm}$ ($Q \approx 2364$), while the passive cavity resonance without monolayer WS$_2$ is $\lambda = 634.1~\text{nm}$ and $\Delta\lambda = 0.1~\text{nm}$ ($Q\approx 6341$). The difference of the measured resonant wavelength to the simulation result could be caused by factors such as the size deviation from fabrication, the presence of monolayer WS$_2$ and possible residue from the transfer process which are not modeled in the simulation. Apart from the dominating cavity peak, a few broad resonant features can be seen modulating the WS$_2$ exciton peak centered around $620~\text{nm}$. Those features are from higher order bulk PhC modes that are not confined by the PhC mode gap. The enhancement from the bulk modes, together with the suspension of monolayer WS$_2$ over the air holes, contributes to the overall enhancement of the on-cavity WS$_2$ exciton emission compared to the off-cavity case. One may notice that the lasing peak is located at the longer wavelength side of the WS$_2$ PL spectrum, which could be due to the reabsorption of WS$_2$ prevents the appearance of gain on the shorter wavelength side\cite{ye2015monolayer,shang2017room}.

\begin{figure}[htb]
 \centering
 \includegraphics[width=\textwidth,keepaspectratio=true]{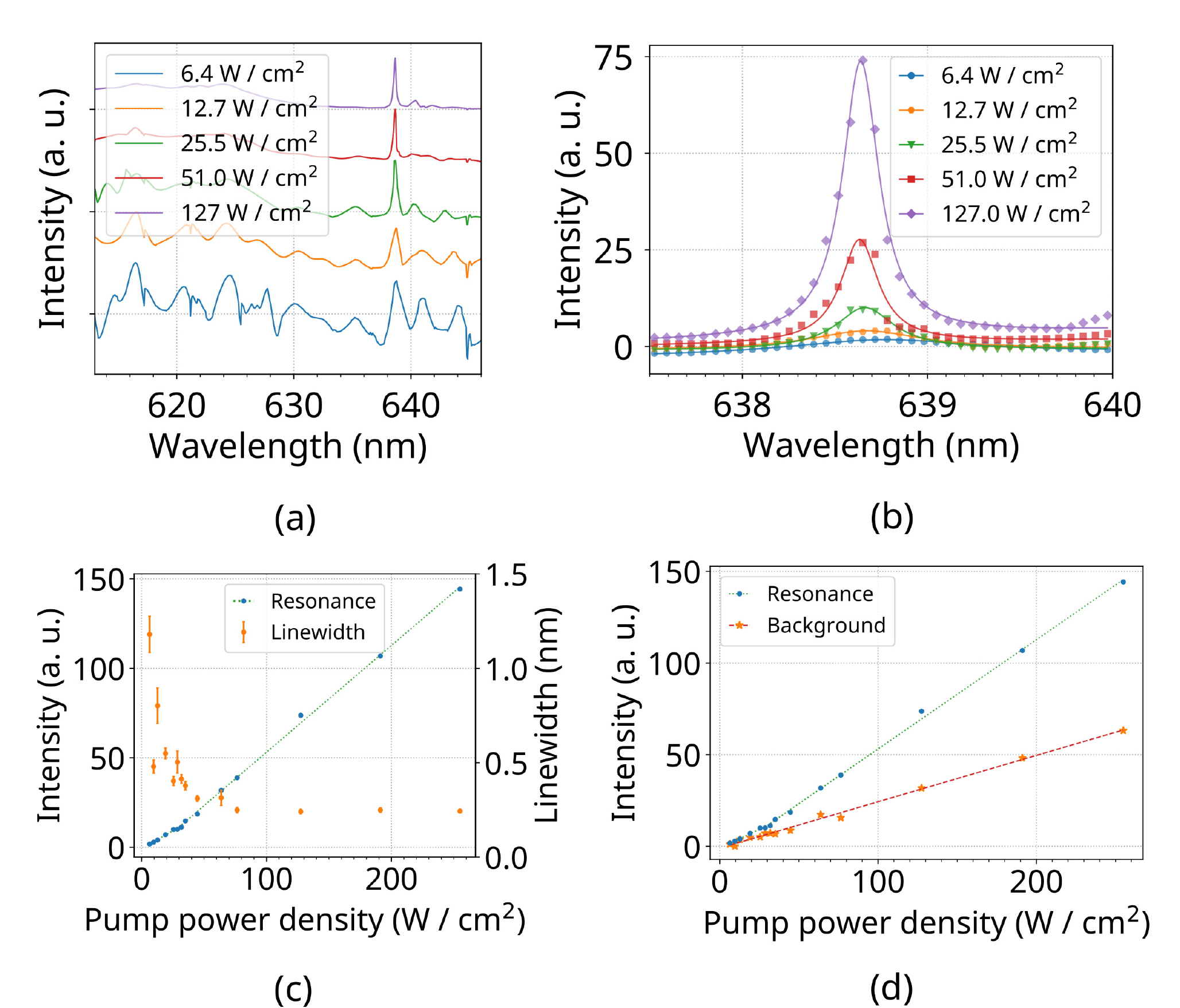}
 \caption{Characterization of the lasing device. (a) Normalized power dependent PL spectra shifted in the vertical direction. (b) Lorentzian fitting of the cavity resonance peak. (c) Resonance peak intensities and fitted linewidths at increasing pump powers. (d) Comparison of the L-L curves of the cavity resonance and the $620~\text{nm}$ background spontaneous emission. The dashed lines in (c) and (d) are the L-L data points fitted with piecewise linear (resonance) or linear (background) functions.}
 \label{fig:pwr}
\end{figure}

To study the transition from spontaneous emission to lasing, we measured the PL spectra of the device at lower pump powers, which are plotted in Fig.~\ref{fig:pwr}(a) with the curves normalized and vertically shifted for clarity. It can be seen that the resonant peak emerges at lower power and eventually outgrows the background emission with increasing pump power, during which narrowing of the linewidth can also be observed. To obtain the exact peak intensities and linewidths, we fitted the peak with a Lorentzian function, as shown in Fig.~\ref{fig:pwr}(b). With those the Light-in, Light-out (L-L) curve of the cavity mode is plotted together with the linewidth evolution in Fig.~\ref{fig:pwr}(c). Profound narrowing of the linewidth can be observed accompanied by a soft kink in the L-L curve. We fitting the L-L curve with a piecewise linear function to determine the location of the change of slope to be $29~\text{W/cm}^2$. Comparing L-L curve with the spontaneous background emission in Fig.~\ref{fig:pwr}(d) taken from the PL intensities at $620~\text{nm}$, the PL intensity of cavity resonance grows faster above the threshold, which verifies the observation from the power dependent spectra. The superlinear behavior and the narrowing of linewidth indicate that lasing is observed in this device.

\begin{figure}[htb]
 \centering
 \includegraphics[width=\textwidth,keepaspectratio=true]{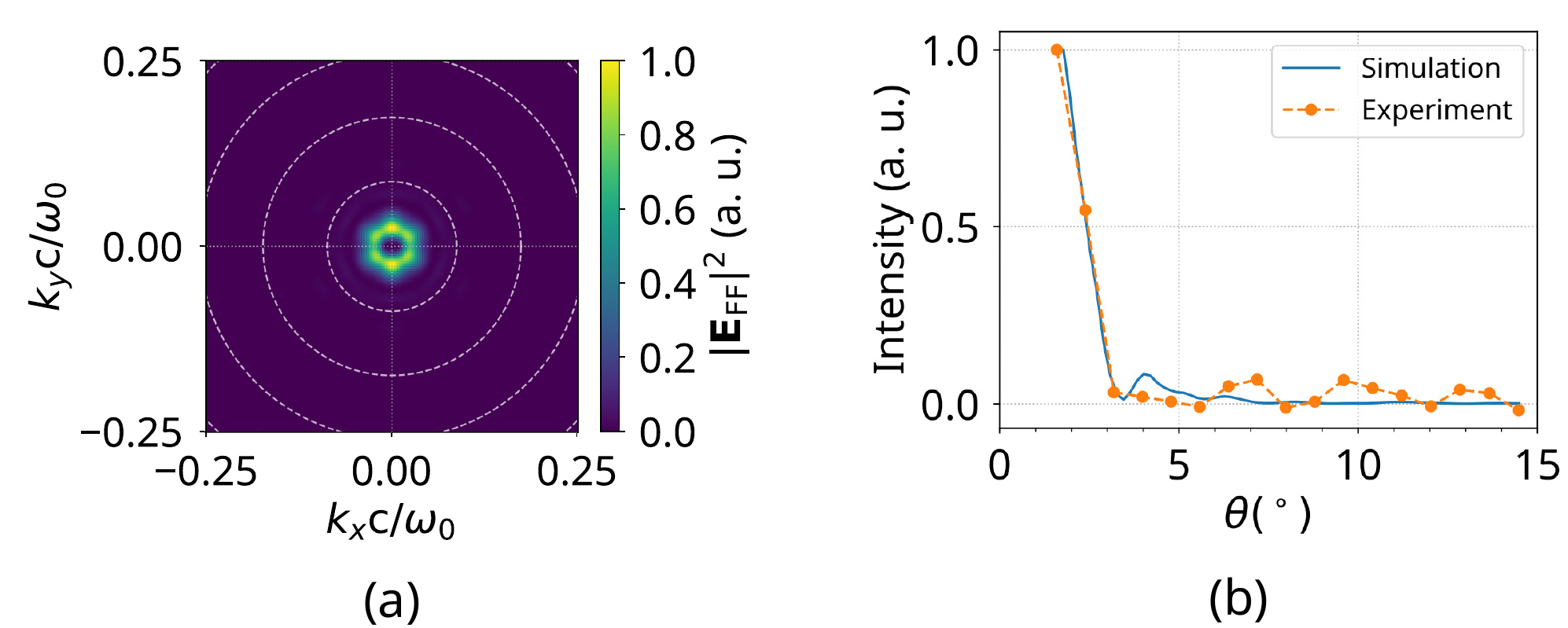}
 \caption{Emission directionality of the WS$_2$ PCSEL. (a) FDTD simulated far field radiation pattern. The dashed circles are the collection angle $\theta$ in $5^\circ$ steps. (b) Comparison between simulated and measured emission profile depending on $\theta$. The minimum angle is $\theta_\text{min} = 1.59^\circ$ for both cases.}
 \label{fig:ff}
\end{figure}

We now turn to the far field property of the laser, as surface normal emission can be expected from such a PCSEL based design. The FDTD simulated far field emission profile of the lasing mode is plotted in Fig.~\ref{fig:ff}(a). Indeed the emission is concentrated near surface normal within a few degrees. To verify that experimentally, the diameter of the iris diaphragm $D$ was used to control the collection angle $\theta(D) \approx \arcsin(D / 2f_\text{obj})$, where $f_\text{obj} = 18~\text{mm}$ is the focal length of the objective lens. Limited by the minimum diameter of the iris diaphragm and the NA of the objective lens, $D$ was changed from $D_\text{min} = 1.0~\text{mm}$ to $D_\text{max} = 9.0~\text{mm}$ in $0.5~\text{mm}$ steps, corresponding to the $\theta$ range from $1.59^\circ$ to $14.48^\circ$. At each $\theta$ angle, the PL spectrum of the cavity pumped above the threshold ($1.27~\text{kW/cm}^2$) was measured and the lasing peak height was extracted by Lorentzian fitting. The angle dependent emission profile can thus be reconstructed by performing a central difference of the accumulated peak heights in respect to the angle $\theta$\cite{nazirizadeh2008optical}, which is plotted in Fig.~\ref{fig:ff}(b). The simulated emission profile is obtained from Fig.~\ref{fig:ff}(a) by a similar treatment. The measured and simulated profiles agree well within the range of $\theta$ measured, indicating that the laser emission from the PhC cavity is concentrated within $\pm 3^\circ$ of the surface-normal direction.



In conclusion, we have demonstrated a room temperature heterostructure photonic crystal surface emitting laser using monolayer WS$_2$ as the gain medium. The near surface-normal far field profile is experimentally measured. We believe that engineering the light extraction, together with shrinking the volume of active region would be a promising approach for realizing low power consumption, high efficiency semiconductor lasers.

\section*{Acknowledgement}
\begin{figure}[htb]
 \centering
 \includegraphics[width=\textwidth,keepaspectratio=true]{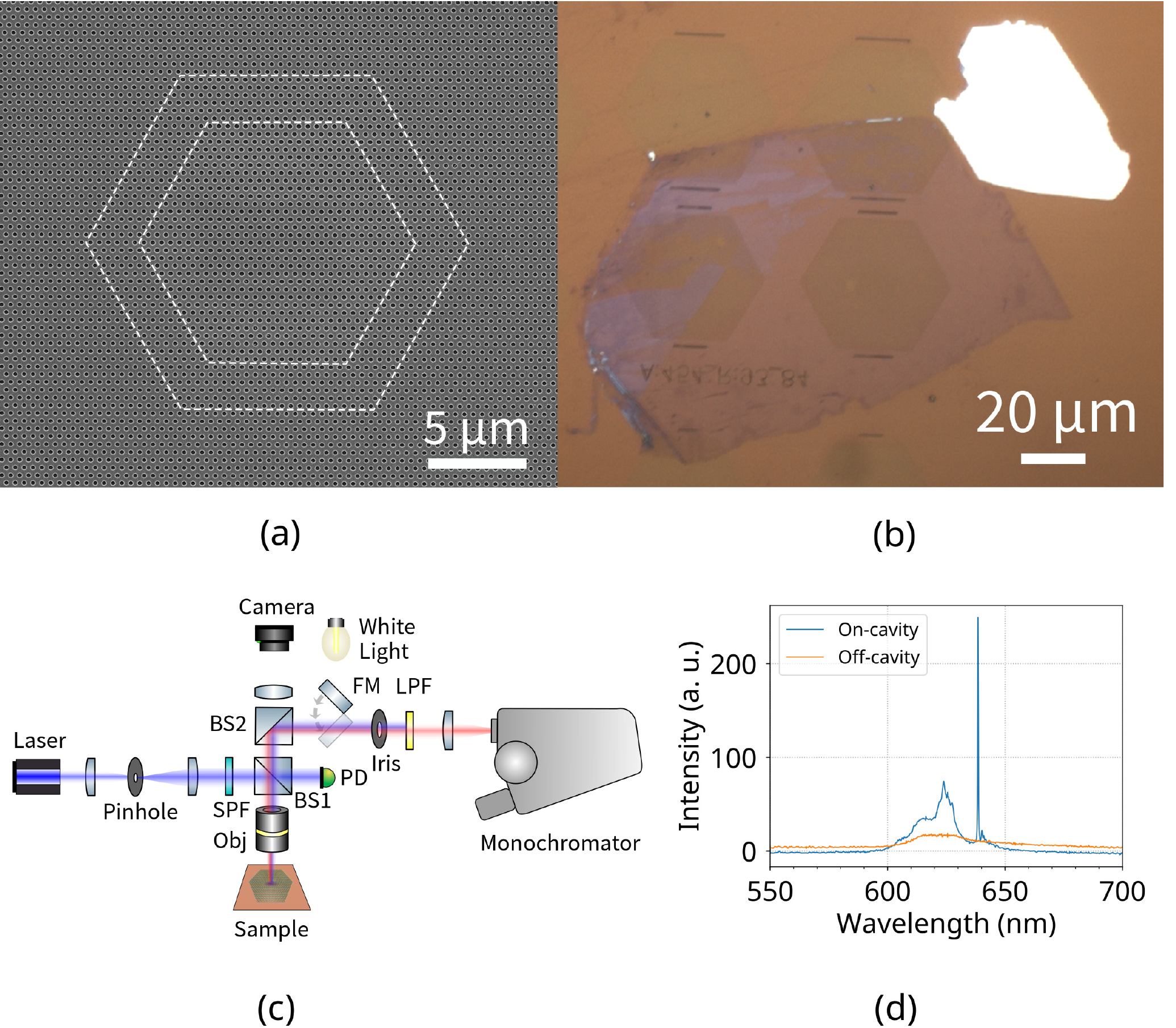}
 \caption{Experimental realization of the WS$_2$ PCSEL. (a) SEM image of a fabricated Si$_3$N$_4$ cavity without WS$_2$. The dashed lines separate PhC regions with different air hole radii. (b) Microscopy image of a WS$_2$ flake transferred onto the Si$_3$N$_4$ PhC cavity. (c) A schematic diagram of the measurement setup. The abbreviations are as follows: SPF: $550~\text{nm}$ short pass filter; BS1 and BS2: Beamsplitters with $R/T$ ratio of $50/50$ and $90/10$ respectively; Obj: $10\times, 0.25~\text{NA}$ objective lens; FM: Flip mirror; PD: Powermeter for monitoring the pump power; LPF: $550~\text{nm}$ long pass filter. (d) Measured PL spectra at $1.27~\text{kW} / \text{cm}^2$ pump power density.}
 \label{fig:exp}
\end{figure}

The presented work is supported by Air Force Office of Scientific Research Grant No. FA9550-16-1-0010 (PM: Dr. Gernot Pomrenke). M. M. acknowledges the funding through the Swiss National Science Foundation project number P300P2\textunderscore 177721. Graphics from ComponentLibrary by Alexander Franzen are used in Fig.~\ref{fig:exp}(c).

\bibliographystyle{unsrtnat}
\bibliography{ref}


\end{document}